\documentclass[aps,prl,footinbib,floats,twocolumn,showpacs,superscriptaddress,longbibliography]{revtex4-1}

\usepackage{natbib}
\usepackage[utf8]{inputenc}
\usepackage{amsmath, amsthm, amssymb}
\usepackage{enumitem}
\usepackage{graphics,graphicx}
\usepackage{times}
\usepackage{dcolumn}
\usepackage{url}
\usepackage{color}

\usepackage[all]{xy}
\usepackage{mdwlist}

\newcommand{\rev}[1]{#1}

\begin{document}

\renewcommand*\thesection{\arabic{section}}
\newcommand{\beq}{\begin{equation}}
\newcommand{\eeq}{\end{equation}}
\newcommand{\sss}{\scriptscriptstyle}

\title{Congestion induced by the structure of multiplex networks}

\author{Albert Sol\'e-Ribalta}
\affiliation{Departament d'Enginyeria Inform\`atica i Matem\`atiques,
Universitat Rovira i Virgili, 43007 Tarragona, Spain}

\author{Sergio G\'omez}
\affiliation{Departament d'Enginyeria Inform\`atica i Matem\`atiques,
Universitat Rovira i Virgili, 43007 Tarragona, Spain}

\author{Alex Arenas}
\affiliation{Departament d'Enginyeria Inform\`atica i Matem\`atiques,
Universitat Rovira i Virgili, 43007 Tarragona, Spain}

\begin{abstract}
Multiplex networks are representations of multilayer interconnected complex networks where the nodes are the same at every layer. They turn out to be good abstractions of the intricate connectivity of multimodal transportation networks, among other types of complex systems. One of the most important critical phenomena arising in such networks is the emergence of congestion in transportation flows. Here we prove analytically that the structure of multiplex networks can induce congestion for flows that otherwise will be decongested if the individual layers were not interconnected. We provide explicit equations for the onset of congestion and approximations that allow to compute this onset from individual descriptors of the individual layers. The observed cooperative phenomenon reminds the Braess' paradox in which adding extra capacity to a network when the moving entities selfishly choose their route can in some cases reduce overall performance. Similarly, in the multiplex structure, the efficiency in transportation can unbalance the transportation loads resulting in unexpected congestion.
\end{abstract}


\maketitle


\section{Introduction}

Complex networks have become a natural abstraction of the interactions between elements in complex systems \cite{Newman2010Book}. When the type of interaction is essentially identical between any two elements, the theory of complex networks provides with a wide set of tools and diagnostics that turn out to be very useful to gain insight in the system under study. However, there are particular cases where this classical approach may lead to misleading results, e.g.\ when the entities under study are related with each other using different types of relations in what is being called multilayer interconnected networks \cite{de2013mathematical,kivela2013multilayer,Boccaletti2014}. Representative examples are multimodal transportation networks \cite{de2014navigability,strano2015multiplex} where two geographic places may be connected by different transport modes, or social networks \cite{mucha2010community,2011MagnaniMultiLayerSocNetwork,granell2013interplay,de2015ranking} where users are connected using several platforms or different categorical layers.

Here, we focus our study on the transportation congestion problem in multiplex networks, where each node is univocally represented in each layer and so the interconnectivity pattern among layers becomes a one-to-one connection
\rev{(i.e., each node in one layer is connected to the same node in the rest of the layers, thus allowing travelling elements to switch layer at all nodes)}.
This representation is an excellent proxy of the structure of multimodal transportation systems in geographic areas \cite{strano2015multiplex}. The particular topology of each layer is conveniently represented as a spatial network where nodes correspond to a certain coarse grain of the common geography at all layers \cite{Barthelemy20111, morris2012transport, louf2013modeling,Louf20140924}.

Transportation dynamics on networks can be, in general, interpreted as the flow of elements from an origin node to a destination node. When the network is facing a number of simultaneous transportation processes, we find that many elements travel through the same node or link. This, in combination with the possible physical constraints of the nodes and links, can lead to network congestion, in which the amount of elements in transit on the network grows proportional to time \cite{guimera2002optimal,zhao2005onset}.
Usually, to analyze the phenomenon, a discrete abstraction of the transportation dynamics in networks is used \cite{tan2014traffic, guimera2002optimal, zhao2005onset, echenique2005dynamics, echenique2004improved, chen2011traffic, cardillo2013modeling}.

Multimodal transportation can also be mathematically abstracted as transportation dynamics on top of a multiplex structure. Note that routings on the multilayer transportation system are substantially different with respect to routings on single layer transportation networks. In the multilayer case, each location of the system (e.g.\ geographical location) has different replicas that represent each entry point to the system using the different transportation media. Thus, each element with the intention of traveling between locations $i$ and $j$ have the option to choose between the most appropriate media to start and end its traversal.
We assume that elements traverse the network using shortest paths, so each element chooses the starting/ending media that minimizes the distance between the starting/ending locations. \rev{As we will show in this work, this ``selfish'' behavior provokes an unbalance in the load of the transportation layers inducing congestion, similarly to what is presented in the classical counterintuitive result of the Braess' paradox \cite{braess}.}

\begin{figure}[!t]
\begin{center}
  \includegraphics[width=0.7\columnwidth,clip=0]{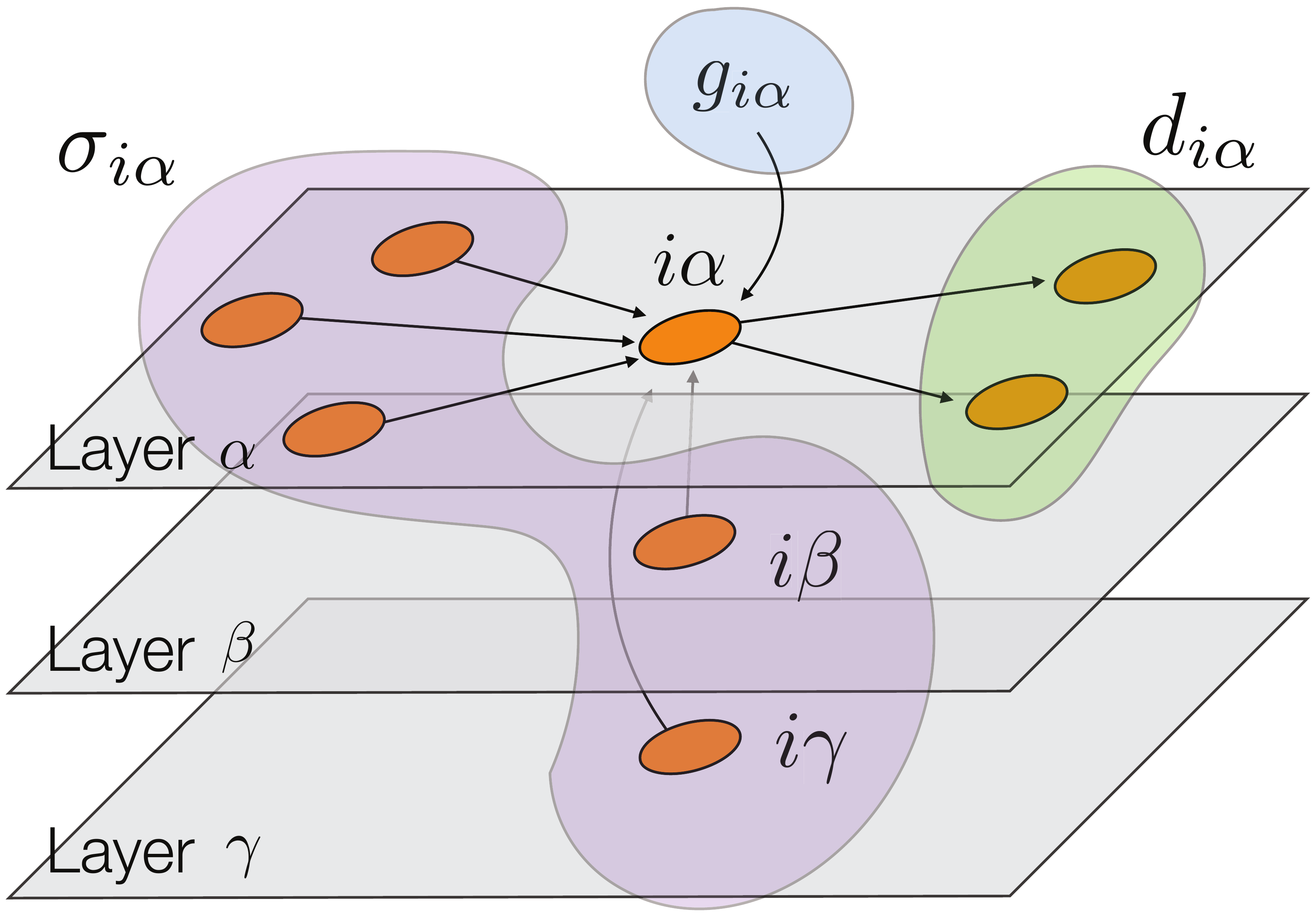}
\end{center}
\caption{Contributions to the size of the queue of each node at each layer of a multiplex network in our standardize transportation model. Arrows mark the flow direction of elements $in$ and $out$ of the node.}
\label{fig:modelExplanation}
\end{figure}

Note that in a multiplex network we can have two types of shortest paths: paths that only use a single layer (intra-layer paths) and paths that use more than one layer (inter-layer paths).
Hereafter, we develop the analysis of transportation in multiplex networks, \rev{consisting of $N$~locations (nodes per layer) and $L$~layers}, and quantify when this structure will induce congestion. To this aim, we describe, with a set of discrete time balance equations, (one for each node at each layer), the increment of elements, $\Delta{q}_{i\alpha}$, in the queue of each node $i$ on layer $\alpha$:
\begin{equation}
	\Delta{q}_{i\alpha} = g_{i\alpha} + \sigma_{i\alpha} - d_{i\alpha},
	\label{eq:balance_equations}
\end{equation}
where $g_{i\alpha}$ is the average number of elements injected at node $i$ in layer $\alpha$ (also called the injection rate, \rev{which can be assimilated to an external particle reservoir}), $\sigma_{i\alpha}$ is the average number of elements that arrive to node $i$ in layer $\alpha$ from the adjacent links of that node (ingoing rate), and $d_{i\alpha} \in [0,\tau_{i\alpha}]$ corresponds to the average number of elements that finish their traversal in node $i$ in layer $\alpha$ or that they are forwarded to other neighboring nodes.  The control parameter is $g_{i\alpha}$: small values of it correspond to low density of elements in the network and high values to high density of elements. A graphical explanation of the variables of the model is shown in Fig.~\ref{fig:modelExplanation}.

Before reaching congestion, the amount of elements in the queue of each node is constant in average, $\Delta{q}_{i\alpha} = 0~\forall i \alpha$ and consequently, $d_{i\alpha} = g_{i\alpha} + \sigma_{i\alpha} < \tau_{i\alpha}$, where $\tau_{i\alpha}$ is the maximum processing rate of the node.  A node $i$ on layer $\alpha$ becomes congested when it is requested to process more elements than its maximum processing rate, $d_{i\alpha} > \tau_{i\alpha}$, and therefore, its onset of congestion is achieved when $d_{i\alpha} = \tau_{i\alpha}$. We are interested on computing the maximum injection rate $g_{i\alpha}$ for which the network is congestion free. In the non-congested phase, as well as on the onset of congestion, the amount of ingoing elements to each node $\sigma_{i\alpha}$ can be obtained in terms of the node's effective betweenness, see \cite{guimera2002optimal}. Our scenario is slightly different since we need to account for the effective betweenness of the multiplex. In addition to the intra-layer and inter-layer paths, our definition of the dynamics also accounts for the \rev{number} of shortest paths that start ($s_{i\alpha}$) and end ($e_{i\alpha}$) at node $i$ on layer $\alpha$
\rev{(this can be computed using any classical shortest path algorithm
\footnote{\rev{In the case of shortest paths degeneracy (multiple shortest paths between the same two locations), their fractional contributions must be considered. For an efficient computation a modifed version of the Brandes' algorithm might be used \cite{brandes2001faster,Sole2014CentralityRankings}.}}).}
\rev{Note that $\sum_{\alpha}s_{i\alpha}=\sum_{\alpha}e_{i\alpha}=N-1$.}
These factors are essential to understand the unbalance of loads between layers in the multiplex network, and only depend on the distribution of shortest path in the full structure.

In the following, we assume a constant injection rate, $\sum_{\alpha} g_{i\alpha} = \rho L$, \rev{being $\rho$ the common injection rate at all locations~$i$}. In addition we also suppose, without loss of generality, that the maximum processing rate is the same for all nodes of the multiplex network, $\tau_{i\alpha} =\tau$. These hypothesis simplify the analysis but are not crucial to develop it.

To obtain the critical injection rate of the multiplex, we require expressions for $g_{i\alpha}$ and $\sigma_{i\alpha}$. The injection rate of node $i$ on layer $\alpha$ can be obtained as the product of the amount of elements that enter the network using location $i$, $\rho L$, and the fraction of multiplex shortest paths that start on node $i$ on layer $\alpha$, \rev{$s_{i\alpha}/(N-1)$}:
\begin{equation}\label{multiplexGBeforeThreshold}
	g_{i\alpha} = \rho L \frac{s_{i\alpha}}{N-1}.
\end{equation}

The ingoing rate of each node, $\sigma_{i\alpha}$, depends on the fraction of shortest paths that pass through or end in it \cite{guimera2002optimal}. Thus, $\sigma_{i\alpha}$ can be obtained as the number of generated elements overall the network at each time step, $\rho L N$, times the fraction of them that arrive ($e_{i\alpha}$) or traverse it ($B_{i\alpha}$, the topological betweenness):
\begin{equation}\label{multiplexSigmaBeforeThreshold}
	\sigma_{i\alpha} = \rho L \frac{B_{i\alpha}+e_{i\alpha}}{N-1},
\end{equation}
When the network is already congested, Eq.~(\ref{multiplexSigmaBeforeThreshold}) does not generally holds since elements traversing congested paths stack in intermediate nodes resulting in a cascade effect not captured by the betweenness. Therefore, our analysis only covers the onset of congestion and it can not be directly applied to the congested regime.

An efficient algorithm to compute the betweenness on multiplex structures can be found in \cite{Sole2014CentralityRankings} for shortest paths dynamics and in \cite{Sole2015RandomWalkCentrality} for random walk dynamics. The computation of $s_{i\alpha}$ and $e_{i\alpha}$ for shortest paths dynamics can be obtained modifying the previously cited algorithm to account for the amount of paths that reach the source and destination nodes.

\begin{figure}[!t]
    \begin{center}
    \includegraphics[width=0.45\textwidth]{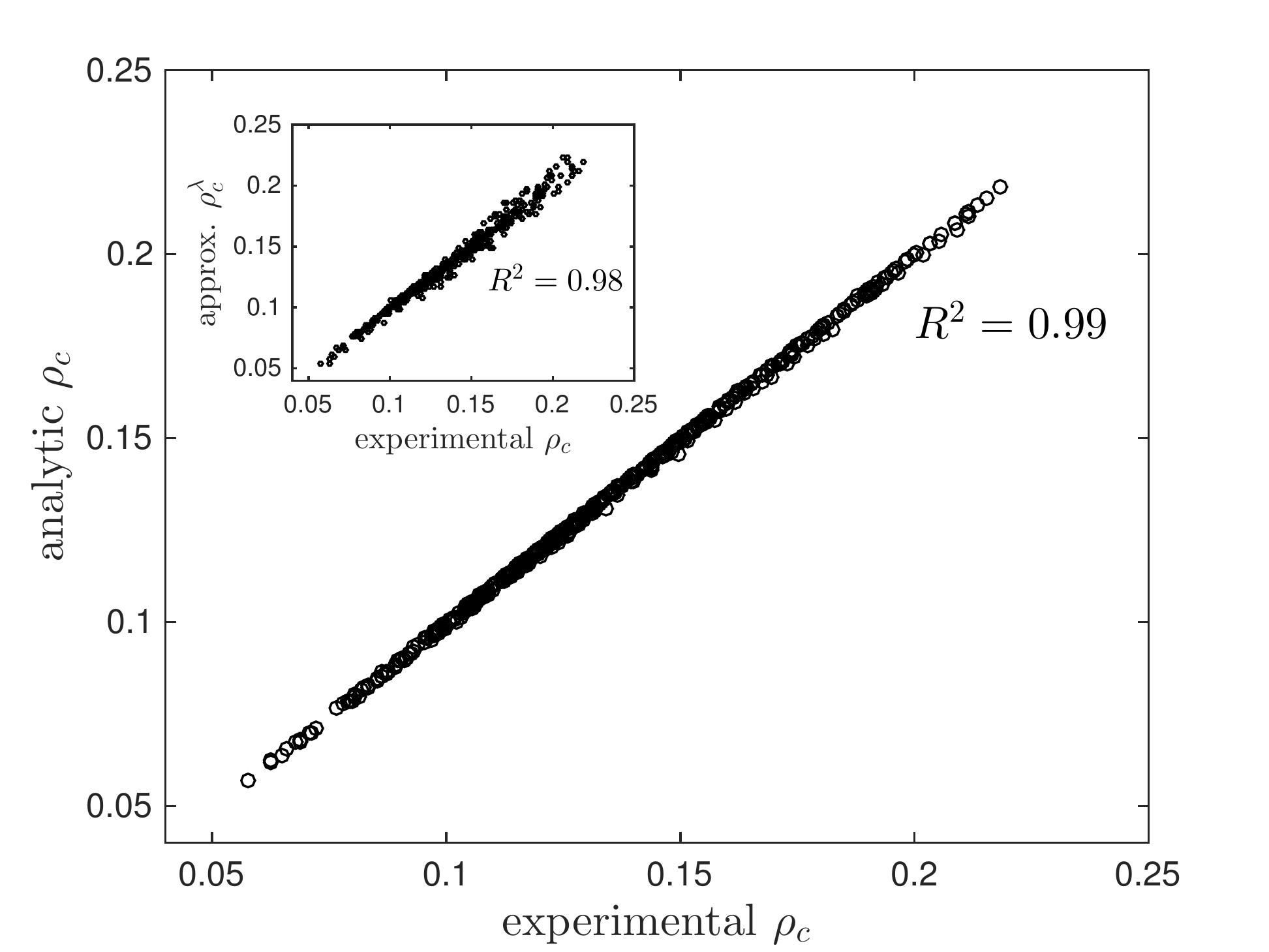}
    \end{center}
        \caption{Accuracy of the analytical value of $\rho_c$ given by Eq.~(\ref{eq:criticalGeneratioRate}) predicting the actual onset of congestion in experimental simulations on 500 random multiplex networks formed by two Erd{\H{o}}s-R\'{e}nyi networks (of 500 nodes) as layers. Inset ({\bf{A}}) shows the correlation between the experimentally obtained critical injection rate and the analytical approximation in Eq.~(\ref{eq:approximatio_rho_c}) where $\lambda$ is approximated by $1$. \rev{$R^2$~is the coefficient of determination for linear fits.}}
        \label{fig:corr_criticGenRate}
\end{figure}

The onset of congestion of the multiplex is attained when a node $i$ in layer $\alpha$ is required to process elements at its maximum processing rate, i.e.\ $g_{i\alpha} + \sigma_{i\alpha} = \tau$. Therefore, the critical injection rate of the system, $\rho_{c}$, becomes
\begin{equation}\label{eq:criticalGeneratioRate}
	\rho_{c} = \tau L^{-1} \frac{N-1}{\mathcal{B}^*},
\end{equation}
where $\mathcal{B}_{i\alpha} \equiv B_{i\alpha}+s_{i\alpha}+e_{i\alpha}$ and $\mathcal{B}^* \equiv \max_{i\alpha} \mathcal{B}_{i\alpha}$. In the following we call $\mathcal{B}_{i\alpha}$ the interconnected betweenness. Note that $\mathcal{B}_{i\alpha}$ depends on intra-layer paths, inter-layer paths, and on the migration of shortest paths between layers (more efficient layers contain a larger proportion of the starting and ending routes). We test the validity of Eq.~(\ref{eq:criticalGeneratioRate}) against Monte Carlo simulations on top of Erd{\H{o}}s-R\'{e}nyi multiplex networks, see Fig.~\ref{fig:corr_criticGenRate}.

In the following, we investigate the role of the topology of the individual layers on the multiplex congestion. First of all, note that  in the definition and computation of the multiplex betweenness (see~\cite{Sole2014CentralityRankings}), the shortest paths (possibly degenerated) between all pair of multiplex locations, $N(N-1)$, are considered. The multiplex structure unbalances, in a highly non-linear way, the distribution of shortest paths among the layers. However, some approximations are possible to grasp the effect of the different contributions to the onset of congestion in multiplex structures.

\begin{figure}[!tb]
    \begin{center}
    \includegraphics[width=0.45\textwidth]{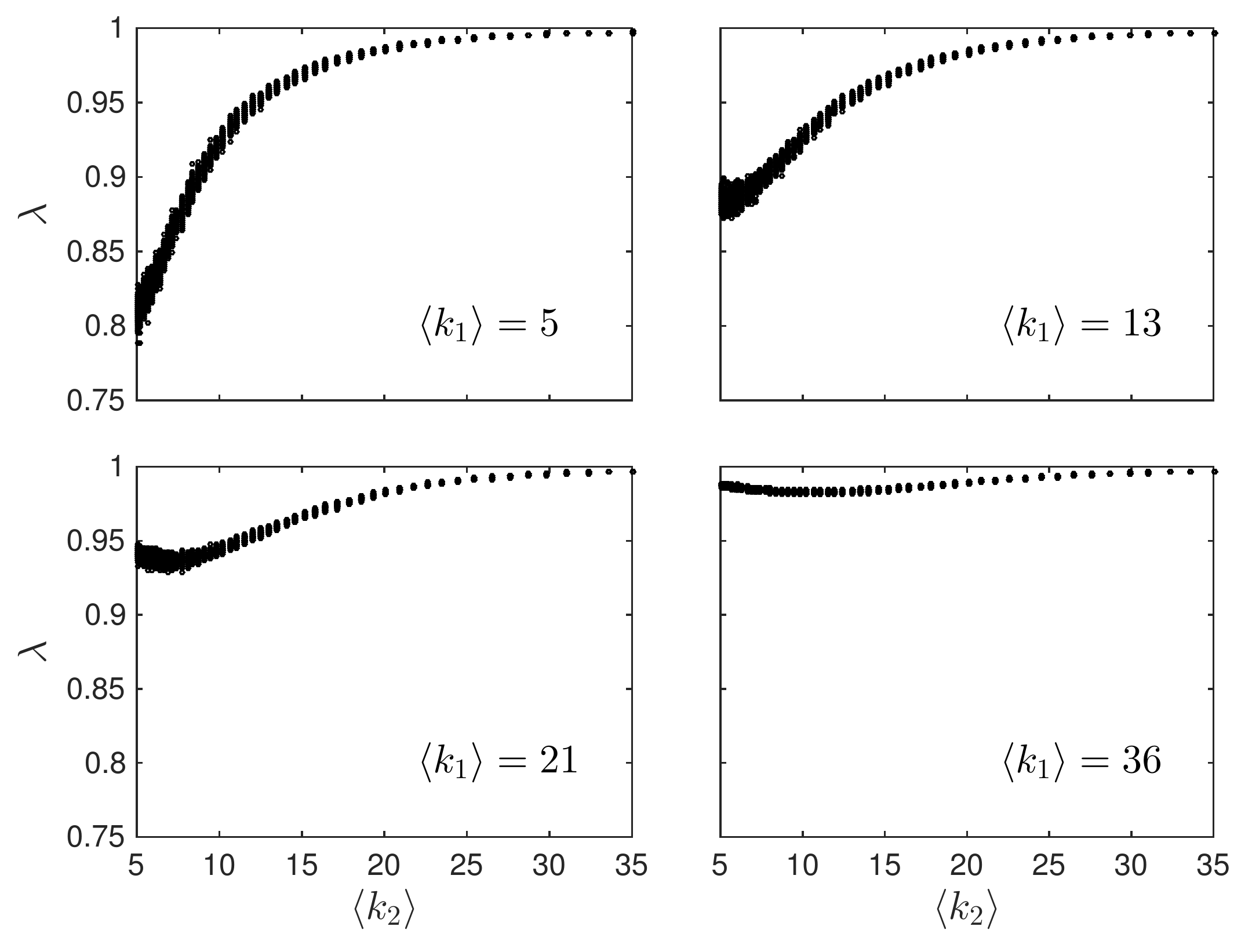}
    \end{center}
        \caption{Fraction $\lambda$ of paths fully contained within layers. Each multiplex network is formed by two Erd{\H{o}}s-R\'{e}nyi layers of 500~nodes each. We plot 100~random realizations for each pair of mean degrees $\langle k_1 \rangle$ and $\langle k_2 \rangle$.}
        \label{fig:percentPathUseMultiplex}
\end{figure}

As stated before, an important parameter of traffic dynamics in multiplex networks is the fraction of inter-layer shortest paths, i.e.\ the fraction of shortest paths that contain, at least, one inter-layer edge. Experiments with multiplex networks composed of two layers, each one being a different random Erd{\H{o}}s-R\'{e}nyi network, show that most of the shortest paths are fully contained within a layer, see Fig.~\ref{fig:percentPathUseMultiplex}. This effect becomes more evident as the degree of the layers increases. Therefore, the fraction of shortest paths fully contained within layers, $\lambda$, is basically 1, and the main factor influencing the traffic dynamics is the migration of shortest paths from the less efficient layer (the one with larger shortest paths) to the most efficient one.
Under this situation we can approximate the interconnected betweenness of node~$i$ in layer~$\alpha$, $\mathcal{B}^{i\alpha}$, in terms of the betweenness of node~$i$ of layer~$\alpha$, $\mathcal{B}_{(\alpha)}^{i}$, when layer~$\alpha$ is considered as a single layer network:
\begin{equation}\label{eq:approx_betweennessMultiplex}
	\mathcal{B}^{i\alpha} \approx \lambda \mu_\alpha \mathcal{B}_{(\alpha)}^{i}\,,
\end{equation}
where $\mu_\alpha<1$ is the fraction of shortest paths using only layer $\alpha$, satisfying $\sum_{\alpha} \mu_\alpha = 1$. The effect of the product of $\lambda \mu_\alpha$ is to precisely account for the fraction of all shortest paths that traverse only layer $\alpha$ in the multiplex. Note that the approximation in Eq.~(\ref{eq:approx_betweennessMultiplex}) does not account for the betweenness contribution of the paths that use inter-layer edges. However, \rev{the high value $\lambda\approx 1$}, indicates that they are usually negligible, and we can even further approximate $\mathcal{B}^{i\alpha} \approx \mu_\alpha \mathcal{B}_{(\alpha)}^{i}$.

Taking advantage of Eq.~(\ref{eq:approx_betweennessMultiplex}), the critical injection rate of the multiplex can be obtained by rescaling the critical injection rate of the individual layers:
\begin{equation}\label{eq:approximatio_rho_c}
	\rho_{c}\approx \tau L^{-1} \frac{N-1}{ \lambda \mu_\ell \mathcal{B^{\ast}_{(\ell)}} } \approx \frac{1}{L\mu_\ell} \rho^{(\ell)}_{c}\,,
\end{equation}
where $\rho^{(\ell)}_{c}$ is the critical injection rate of the most efficient layer~$\ell$. Fractions $\mu_\ell$ and $\lambda$ are genuine properties of the multiplex network structure that can be obtained by means of the multiplex extension of the Brandes' betweenness algorithm \cite{Sole2014CentralityRankings}. Figure~\ref{fig:corr_criticGenRate} {\bf (A)} shows the accuracy of this approximation in the calculation of $\rho_c$.
The high accuracy obtained in the approximation evidences that the critical injection rate of the multiplex crucially depends on the migration of shortest paths between layers, which is captured in $\mu_{\ell}$.

As an example, consider a multiplex structure composed by two identical layers. In this case, there are no shortest paths using inter-layer edges since they would be longer than the ones fully included in one layer, thus $\lambda=1$.  Since paths in both layers are identical, there is a multiplex path degeneration: for each shortest path in layer~1 there is an equivalent shortest path in layer~2. As a consequence, nodes on the paths only obtain $1/2$ of the betweenness contribution they would obtain if layers were separated, which results in $\mu_\ell = 1/2$. Eventually, we see that for identical layers the multiplex betweenness is $1/2$ of the betweenness computed on any of the layers.

\begin{figure*}[t]
    	\begin{tabular}{ll}
                {\bf (A)} & {\bf (B)}
                \\
                \includegraphics[width=0.35\textwidth]{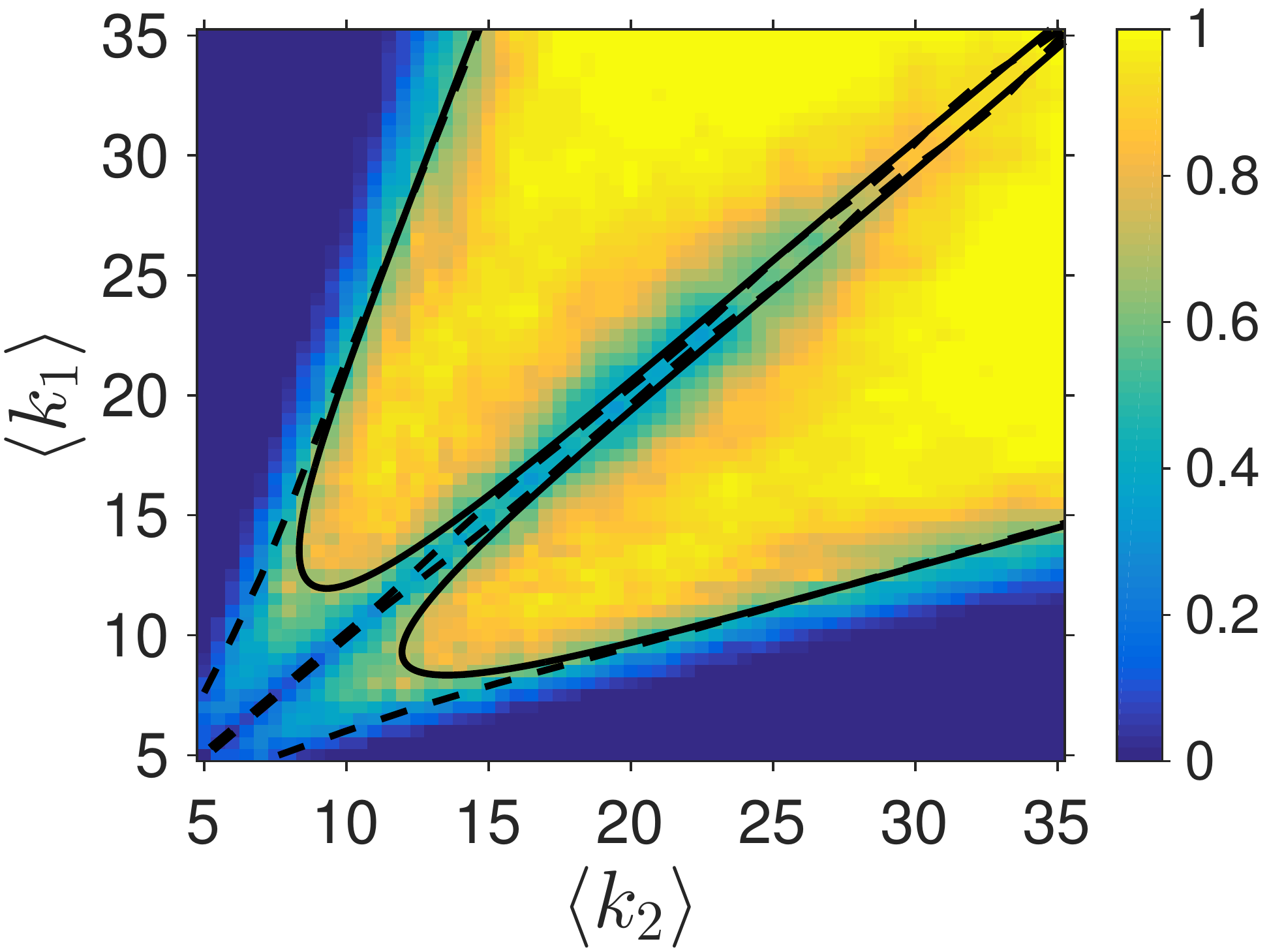}
                &
                \includegraphics[width=0.35\textwidth]{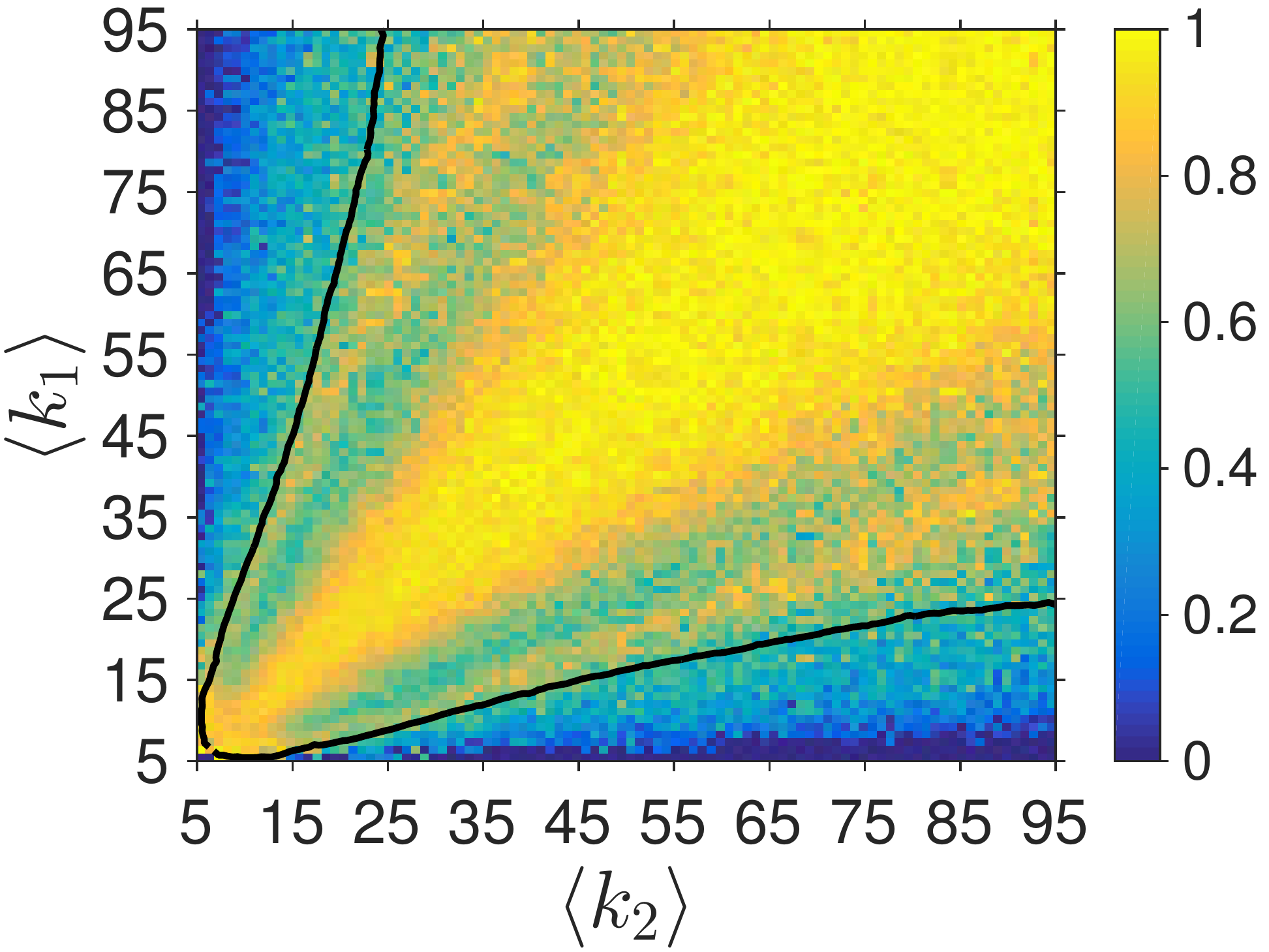}
        \end{tabular}
        \caption{Probability of obtaining a multiplex configuration that induces congestion when: {\bf (A)} the multiplex is composed of two Erd{\H{o}}s-R\'{e}nyi layers; {\bf (B)} the topology is a Random Geometric Multiplex. In both networks topologies each layer has 500 nodes. \rev{The number of simulations points is $50^2$, and for each point we generate $10^2$ configurations fixing $\langle k_1 \rangle$ and $\langle k_2 \rangle$. The colors indicate the probability of observing that the onset of congestion of the multiplex satisfies $\rho_c < \min(\rho_c^{(1)},\rho_c^{(2)})$}.         Lines show the accuracy of Eq.~\ref{eq:congestionInducedByMultiplex_lambdamu} in detecting the region where multiplex structure induces congestion. Solid lines represent the expression when the real value of $\lambda$ is used and dashed lines when we approximate $\lambda$ by $1$.}
        \label{fig:probabilityMoreLessResilient}
\end{figure*}

On the other side, consider a multiplex network in which most of the paths in layer~$1$ have length~$2$ and most of the paths in layer~$2$ have length~$3$. Again, there are very few shortest paths using inter-layer edges since their minimum length is~$3$ (i.e.\ one intra-layer edge, followed by a change of layer through an inter-layer link, and finally another intra-layer edge), therefore $\lambda \approx 1$. Moreover, most of the shortest paths make use of layer~$1$, where the lengths are shorter, so $\mu_1 \approx 1$ and $\mu_2 \approx 0$. Substitution in Eq.~\ref{eq:approx_betweennessMultiplex} shows that the interconnected betweenness of the multiplex is equivalent to the betweenness of the most efficient layer, that in this case is layer~$1$.

We can compute the congestion induced by a multiplex as the situation in which a multiplex network reaches congestion with less load than the worst of its layers when operating individually. In a multiplex with two layers $1$ and $2$ (being $2$ the most efficient), this limiting situation is obtained when $\rho_{c}<\rho_{c}^{(1)}$, and consequently:
\begin{equation}\label{eq:congestionInducedByMultiplex_lambdamu}
	\frac{1}{L\lambda\mu_{2} }  \lesssim \frac{\mathcal{B}^{\ast}_{(2)}}{\mathcal{B}^{\ast}_{(1)}}.
\end{equation}

Figure~\ref{fig:probabilityMoreLessResilient}{\bf (A)} shows the regions where the multiplex structure induces congestion for sets of Erd{\H{o}}s-R\'{e}nyi multiplex networks. In each experiment, two Erd{\H{o}}s-R\'{e}nyi networks with different mean degree are coupled to form a multiplex network. For each pair of mean degrees we have evaluated 100~random realizations of the multiplex network and for each realization we have computed the onset of congestion of the multiplex network and of the individual layers. We have then obtained the fraction of times that the multiplex network reaches congestion before both layers. The boundaries approximated by Equation~\ref{eq:congestionInducedByMultiplex_lambdamu} determine accurately the regions where the multiplex induces congestion. As expected, the approximation using only $\mu$ works well except when both mean degrees are low since on these cases the amount of shortest paths using the multiplex structure is more relevant. \rev{Surprisingly, for larger degrees (in the diagonal) the ER networks generated present small fluctuations on the average degree that eventually make a node in one layer to have a maximum degree a little bit larger than in the other layer. This asymmetry, for such dense networks, is enough to provoke a load unbalance that is reflected in the simulations.}

\begin{figure}[t]
    \begin{center}
    \includegraphics[width=0.35\textwidth]{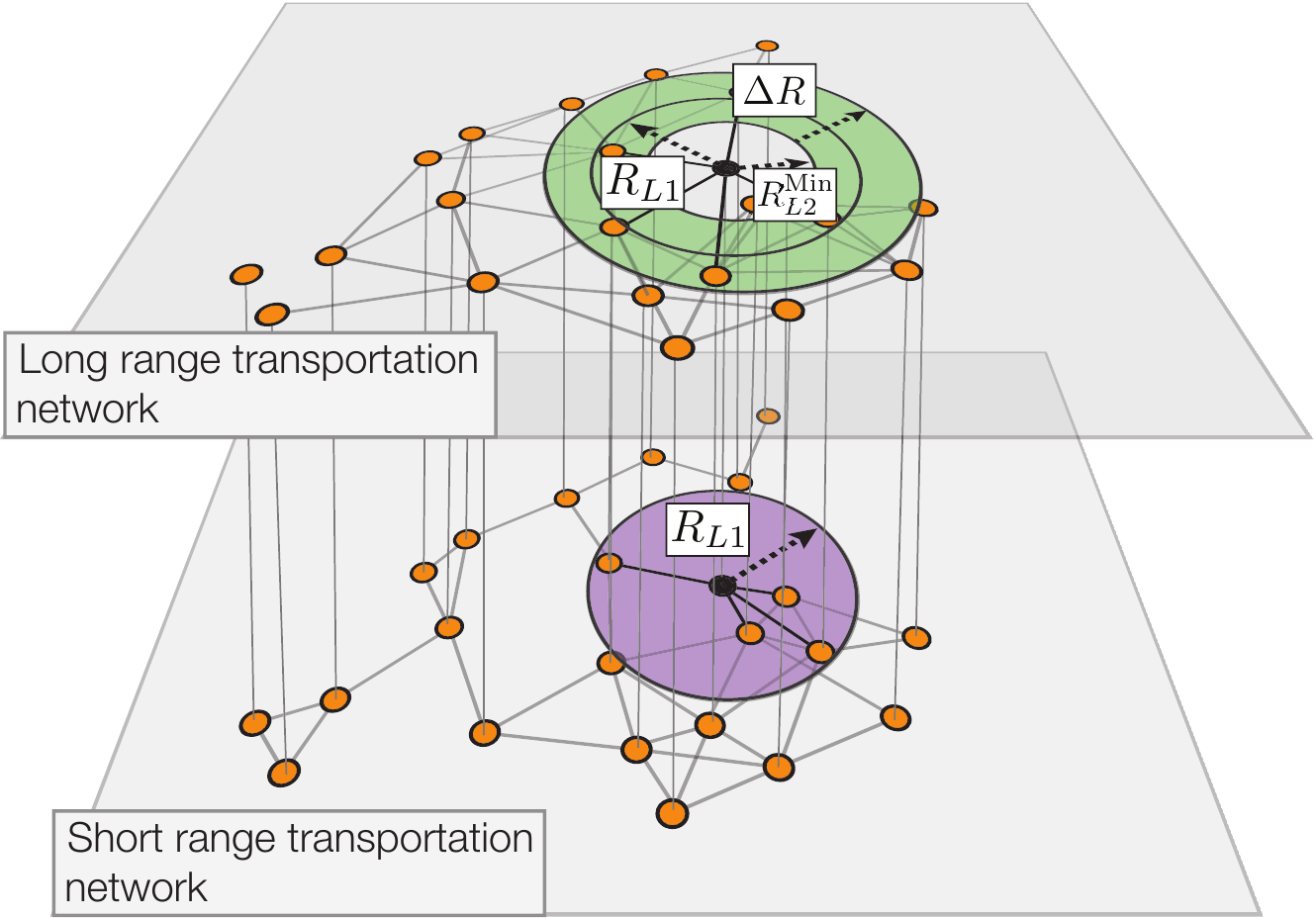}
    \end{center}
        \caption{Sketch of the generation process of a random geometric multiplex. We first choose uniformly at random a set of N points in a bi-dimensional space, $(x,y) \in [0,1]^2$; these are our node locations. We then generate the first layer by adding edges between all locations $i$ and $j$ separated by a distance $d_{ij}$ lower than a certain radius $R_{L1}\in [0,0.4]$. The second layer is generated by adding edges between all node pairs with distance $R^{\mbox{\scriptsize Min}}_{L2}<d_{ij}<R^{\mbox{\scriptsize Max}}_{L2}$. The values of $R^{\mbox{\scriptsize Min}}_{L2} \in [0,R_{L1}]$ force minimum overlapping between both layers.  The value of $R^{\mbox{\scriptsize Max}}_{L2} = R^{\mbox{\scriptsize Min}}_{L2} + \Delta R$ with $\Delta R \in [0,R_{L1}]$ ensures the range $[R^{\mbox{\scriptsize Min}}_{L2},R^{\mbox{\scriptsize Max}}_{L2}]$ does not exceed the radius of first layer.}
        \label{fig:geometricMultiplex}
\end{figure}

We have used homogenous random networks multiplexes to demonstrate the use of the analytical approach, however the theory is general for any other multiplex network structure. To conclude this letter, we have also used a different type of topology, random geometric graphs, more akin to represent transportation networks \cite{taylor2015topological,Barthelemy20111}. To this end, we propose a simple configuration of a random geometric multiplex. We assume each random geometric multiplex is composed of two types of transportation media: short range (e.g.\ the bus network) and \rev{long range} (e.g.\ the subway), see Fig.~\ref{fig:geometricMultiplex}.

Our construction method allows to generate very extreme geometric multiplexes; from configurations where the \rev{long range} layer only contains some of the longer edges of the short range layer ($R^{\mbox{\scriptsize Max}}_{L2}\approx R_{L1}$), to a \rev{long range} layer that only contains edges larger than the ones in the short range layer ($R^{\mbox{\scriptsize Min}}_{L2}\approx R_{L1}$). However, we usually obtain configurations where the \rev{long range} layer have some degree of edge overlap with the short range layer. The test set where we have performed the experiments has been constructed by creating $10^5$ random geometric multiplex networks choosing uniformly at random the parameters of the model. Figure~\ref{fig:probabilityMoreLessResilient}{\bf (B)} shows that Eq.~\ref{eq:congestionInducedByMultiplex_lambdamu} accurately predicts the region where the multiplex structure induces congestion.

In summary, we have analyzed the congestion phenomena on multiplex transportation networks. We have developed an standardized model of how elements traverse those networks and we have provided analytical expression for the onset of congestion. Then, we have shown that the multiplex structure induces congestion and derived analytical expressions to determine the network parameters that raise this phenomena. All analytical expressions have been assessed on Erd{\H{o}}s-R\'{e}nyi and Geometric multiplex networks, and showing a perfect agreement with the empirical results. \rev{The reason behind this phenomenology is the unbalance of shortest paths between layers. The flow follows the shortest path, increasing the load of the most efficient (in terms of shortest paths) layer, and eventually congesting it.}
Theory and experiments developed in this paper are specially useful to understand transportation dynamics on multilayer networks and might help on the development of more efficient transportation networks and routing algorithms.

\section{Acknowledgements}

This work has been supported by Ministerio de Econom\'{\i}a y Competitividad (Grant FIS2012-38266) and European Comission FET-Proactive Projects PLEXMATH (Grant 317614). A.A.~also acknowledges partial financial support from the ICREA Academia and the James S. McDonnell Foundation.

\end{document}